\documentstyle[12pt]{article}
\begin {document}
\title{Isgur-Wise functions from the MIT bag \\ model
\thanks{Work supported in part by the KBN grants no.\ 20-38-09101.}
}
\author{
{\sc M.\,Sadzikowski}\thanks{E-mail {\sc UFSADZIK@PLKRCY11.BITNET}} \\
{\it Institute of Physics, Jagellonian University,} \\
{\it ul.\,Reymonta 4, PL-30\,059 Krak\'ow, Poland.}
\vspace{0.5cm}\\
and
\vspace{0.5cm}\\
{\sc K.\,Zalewski}\thanks{E-mail {\sc ZALEWSKI@VSK01.IFJ.EDU.PL}}\,\thanks{
also Institut of Nuclear Physics, Krak\'ow}\\
{\it Institute of Physics, Jagellonian University,} \\
{\it ul.\,Reymonta 4, PL-30\,059 Krak\'ow, Poland.}}
\date{}

\maketitle
\begin{center}

\begin{abstract}
The Isgur-Wise functions for the ground state to ground state semileptonic
decays involving $b \rightarrow c$ transitions are calculated from the
(modified) MIT bag model. It is checked that the results for the decays
$\overline{B} \rightarrow D l \overline\nu$ and $\overline{B} \rightarrow D^*
l \overline\nu$ agree well with experiment. Predictions for the decays
$\Lambda_b \rightarrow \Lambda_c l \overline\nu$, $\overline{B}_s \rightarrow
D_s l \overline\nu$ and $\overline{B}_s \rightarrow D^*_s l \overline\nu$ are
given and discussed.

\end{abstract}
\end{center}

\section{Introduction}

According to the heavy quark paradigm a hadron containing one heavy quark
consists of the heavy quark almost immobile at the centre of the hadron and of
a light component surrounding it. In general the light component is very
complicated. In the valence quark approximation, however, the light component
of a meson is a light (anti)quark and the light component of a baryon is a
light diquark. This picture, familiar from the nonrelativistic quark model, cf.
e.g. \cite{ISGW}, can be derived in a model independent way from QCD, when the
mass of the heavy quark $m_Q$ tends to infinity \cite{IW1}. Therefore, each
good model retaining its validity in the limit $m_Q \rightarrow \infty$ should
be possible to interpret as the heavy quark picture modified by some finite
mass corrections.

One of us (M.S.) extending and updating ealier work \cite{PON}, \cite{SHU},
\cite{ITS} has discussed from this point of view the predictions of a suitably
modified MIT bag model for the mass spectra of mesons and baryons containing
heavy quarks \cite{SAD}. In the present paper we discuss some semileptonic
decays, where the $b$ quark goes over into a $c$ quark and an $l\overline\nu$
pair. Our main result is that the bag model gives very reasonable Isgur-Wise
functions. For the decays $\overline{B} \rightarrow Dl\overline\nu$ and
$\overline{B} \rightarrow D^*l\overline\nu$ (in the following denoted together
$\overline{B} \rightarrow D^{(*)}l\overline\nu$ ) these results can be directly
compared with experiment and the agreement is very satisfactory. We also give
predictions for the related decay $\overline{B}_s \rightarrow D^{(*)}_s
l\overline\nu$. For the decays $\Lambda_b \rightarrow \Lambda_c l\overline\nu$
we find that the Isgur-Wise function is much steeper than for $B$ decays and
consequently we predict that the fraction of the decays $\Lambda_b \rightarrow
X_c l\overline\nu$, where $X_s = \Lambda_c$ is smaller than the corresponding
fraction for the decays $\overline{B} \rightarrow D^{(*)}l\overline\nu$ in
spite of kinematical factors working in the opposite direction. Moreover, we
derive a new, restrictive and almost model-independent lower bound for the
parameters $\rho^2$ related to the Isgur-Wise functions. Finally, we comment on
the implications of our analysis for rare decays like $B \rightarrow
K^*\gamma$.

\section{Decays $H_b \rightarrow H_c l\overline\nu$  in the heavy quark limit}

The standard formula \cite{PDG} for the width of the decay of hadron $H_b$
into hadron $H_c$ and an $l\overline\nu$ pair can be written in the form

\begin{equation}
\label{decwid}
\frac{d\Gamma}{dq^2} =
\frac{|\vec{p}_{Hc}|}{128\pi^3m_{Hb}^2}\overline{|{\cal{M}}|^2}
\end{equation}

\noindent The momentum of $H_c$ is calculated in the $H_b$ rest frame and the
square of the absolute value of the invariant amplitude $\cal{M}$ is averaged
over the angular distribution of $H_c$ in the $H_b$ rest frame and of the
lepton in the $l\overline\nu$ rest frame. We define $q^\mu = p_l^\mu +
p_{\overline\nu}^\mu$ so that the $l\overline\nu$ rest frame can be referred to
as the $\vec{q} = 0$ frame. The invariant amplitude can be written in the form

\begin{equation}
\label{amplit}
{\cal{M}} = \frac{G_FV_{cb}}{\sqrt{2}}\overline{u}_e\gamma^\mu(1 -
\gamma^5)v_\nu
<H_c|J_\mu|H_b>
\end{equation}

\noindent We choose the $z$-axis in the $\vec{q} = 0$ frame along the momentum
of $H_b$ as seen in this frame. In the $\vec{q}=0$ frame, neglecting the masses
of the leptons, the time-component of the leptonic current vanishes and the
space components are

\begin{equation}
2\overline{u}_e\vec{\gamma}v_\nu = -2\sqrt{q^2}(i\sin\phi_l +
\cos\theta_l\cos\phi_l, -i\cos\phi_l + \cos\theta_l\sin\phi_l, -\sin\theta_l)
\end{equation}

In the heavy quark limit (further quoted HQET) the hadron current can be
expressed by the quark current and the Isgur-Wise functions. The formula
\cite{KZ1} reads

\begin{eqnarray}
\label{clegor}
<H_c|J_\mu|H_b> = \sum_{\lambda}
<J_{Hbl},\lambda;\frac{1}{2},\lambda_b|J_{Hb},\lambda_{Hb}> \nonumber \\
<J_{Hcl},\lambda;\frac{1}{2},\lambda_c|J_{Hc},\lambda_{Hc}>
\sqrt{\frac{\omega+1}{2}}
\overline{u}_c\gamma_\mu(1-\gamma^5)u_b\xi_\lambda(\omega)
\end{eqnarray}

\noindent Here $<\ldots|\ldots>$ are ordinary Clebsch Gordan coefficients,
$J_{HQl}$ is the angular momentum of the light component of hadron $H_Q$ in the
$H_Q$ rest frame and $\lambda$ is its (conserved) projection on the $z$-axis
chosen in the $H_b$ rest frame along the direction of the $H_c$ momentum. The
Isgur-Wise functions $\xi_\lambda(\omega)$ are, except for a trivial
kinematical factor, overlap functions of the light component of $H_b$ at rest
with the light component of $H_c$ moving with the Lorentz factor $\omega$ in
the $+z$ direction

\begin{equation}
\xi_\lambda(\omega) = \sqrt{\frac{2}{\omega+1}}<\Phi_{Hcl}|\Phi_{Hbl}>
\end{equation}

\noindent Since the Isgur-Wise function depends only on the light components,
it does not depend on the spin $J_{HQ}$. For instance, it is the same for both
decays $\overline{B} \rightarrow D^{(*)} l\overline\nu$.

The number of independent Isgur-Wise functions for a given decay is reduced by
the identity \cite{KZ1}

\begin{equation}
\xi_\lambda(\omega) = \eta_{Hb}\eta_{Hc}(-1)^{J_{Hc} -
J_{Hb}}\xi_{-\lambda}(\omega)
\end{equation}

\noindent where $\eta_{HQ}$ is the internal parity of the light component of
hadron $H_Q$.

The quark current in the $H_b$ rest frame is

\begin{equation}
\overline{u}_c\gamma^\mu(1-\gamma^5)u_b =
\sqrt{m_bm_c}\sqrt{\frac{2}{\omega+1}}(\omega + 1 -2\lambda_cv_c)\phi^{\dag}_c
\tilde{\sigma}^\mu\phi_b
\end{equation}

\noindent In this formula: $v_c = |\vec{v}_c|$. Here and in the following by
velocity we always mean the fourvector velocity $v^\mu = p^\mu/m$. The quark
spinors are $\phi_i = \left(\begin{array}{c}1\\0\end{array}\right)$ for
helicity $+1/2$ and $\phi_i = \left(\begin{array}{c}0\\1\end{array}\right)$ for
helicty $-1/2$. The object $\tilde\sigma^\mu = (1,-\vec{\sigma})$. Substituting
this current into formula (\ref{clegor}) and contracting the result with the
lepton current transformed to the $H_b$ rest frame:

\begin{eqnarray}
\overline{u}_l\gamma^\mu(1-\gamma^5)v_\nu = 2(q_z\sin{\theta_l},
\sqrt{q^2}(i\sin{\phi_l}+\cos{\theta_l}\cos{\phi_l}),\nonumber \\
\sqrt{q^2}(-i\cos{\phi_l}+\cos{\theta_l}\sin{\phi_l}), q_0\sin{\theta_l})
\end{eqnarray}

\noindent one finds from (\ref{amplit}) the invariant amplitude and from
(\ref{decwid}) the decay width.

For further reference we quote the results for the decay $\overline{B}
\rightarrow Dl\overline\nu$:

\begin{equation}
\frac{d\Gamma}{dq^2} =  \frac{G_F^2|V_{cb}|^2|\vec{p}_D|^3}{24\pi^3}
\frac{(m_B+m_D)^2}{4m_Bm_D}\xi^2_{\frac{1}{2}}(\omega)
\end{equation}

\noindent for $\overline{B} \rightarrow D^*l\overline\nu$ (the subscripts of
$\Gamma$ refer to the helicities of $D^*$):

\begin{eqnarray}
\frac{d\Gamma_\pm}{dq^2} &=& \frac{G_F^2|V_{cb}|^2|\vec{p}_{D^*}|}{48\pi^3}
\frac{m_{D^*}}{m_B} q^2 (\omega \mp v_{D^*})(\omega+1)
\xi^2_{\frac{1}{2}}(\omega)\\
\frac{d\Gamma_0}{dq^2} &=& \frac{G_F^2|V_{cb}|^2|\vec{p}_{D^*}|}{96\pi^3}
\frac{m_{D^*}}{m_B}(m_B - m_{D^*})^2 (\omega+1)^2 \xi^2_{\frac{1}{2}}(\omega)
\end{eqnarray}

\noindent and for $\Lambda_b \rightarrow \Lambda_c l \overline\nu$ (averaged
over the initial helicity):

\begin{equation}
\frac{d\Gamma}{dq^2} = \frac{G_F^2|V_{cb}|^2|\vec{p}_{\Lambda_c}|}{48\pi^3}
\frac{m_{\Lambda_c}}{m_{\Lambda_b}} [3\omega{}q^2 +
2m_{\Lambda_c}m_{\Lambda_b}(\omega^2-1)](\omega+1)
\overline\xi^2_{\frac{1}{2}}(\omega)
\end{equation}

\noindent The bar over $\xi$ in the last formula serves as a remainder that
this Isgur-Wise function is different from that in the preceding formulae. For
the corresponding decays involving $\overline{b} \rightarrow \overline{c}$
transitions the only modification in the formulae is that the sign of $v_c$
should be changed.

According to custom, we have put everywhere particle masses for the quark
masses. The difference being of order $O(m_Q^0)$ is formally negligible at
leading order of HQET, but numerically it may be quite important. With obvious
changes these formulae can be also applied to related decays like
$\overline{B}_s \rightarrow D^{(*)}_sl\overline\nu$.

It is seen that the only unknown factors in these formulae are the Isgur-Wise
functions $\xi$. Even they drop out from expressions for ratios of decay
widths of a given decay into various helicity states at given $q^2$. For
instance for the decay $\overline{B} \rightarrow D^*l\overline\nu$ the
polarization coefficient $\alpha$ is related to

\begin{equation}
\left( \frac{d\Gamma_+}{dq^2} + \frac{d\Gamma_-
}{dq^2}\right)/2\frac{d\Gamma_0}{dq^2} = \frac{2\omega}{\omega+1}
\frac{q^2}{q_{max}^2}
\end{equation}

\noindent where the maximum square of the momentum transer $q_{max}^2 = (m_B -
m_{D^*})^2$, while the asymmetry parameter $A_{fb}$ is

\begin{equation}
-\frac{3}{4}\left( \frac{d\Gamma_+}{dq^2} - \frac{d\Gamma_- }{dq^2}\right)/
\left( \frac{d\Gamma_+}{dq^2} + \frac{d\Gamma_0}{dq^2} + \frac{d\Gamma_-
}{dq^2}\right) = \frac{3q^2}{4q^2v_{D^*} +(\omega+1)q^2_{max}}v_{D^*}
\end{equation}

\noindent In practice, however, in order to compare these formulae with
experiment one has to average over $q^2$ with weights taking into account the
experimenal cut offs. This again requires a knowledge of the Isgur-Wise
functions.

\section{The Isgur-Wise functions}

The calculation of the Isgur-Wise functions $\xi_\lambda(\omega)$ reduces to
the calculation of the overlaps $<\Phi_{Hcl}|\Phi_{Hbl}>$. In the present paper
we consider only the so-called ground state to ground state transitions, where
$\xi(1) = 1$ \footnote{In the following we suppress the subscript $\lambda$,
since this leads to no confusion}. Starting from a model reliable for small
recoil velocities, one can either argue that the recoils occuring in the
semileptonic $b \rightarrow c$ decays are small (cf. e.g. \cite{ISG}), or
determine the parameter $\rho^2$ in the expansion

\begin{equation}
\xi(\omega) = 1 - \rho^2(\omega - 1) + O((\omega-1)^2)
\end{equation}

\noindent and use some plausible functional form to continue $\xi(\omega)$ (cf.
e.g. \cite{ISGW}). A formula particularly convenient for calculations is
\cite{ISGW}

\begin{equation}
\xi(\omega) = exp[-\rho^2(\omega - 1)]
\end{equation}

\noindent but many other formulae have been suggested \cite{ROS}, \cite{NR},
\cite{RAD}, \cite{NEU}, \cite{DRT}, \cite{MS}, \cite{EFI}. The values of
$\rho^2$ proposed in the literature range from $\frac{1}{3}$ \cite{EFI} to
infinity \cite{RAD}. For given data the best $\rho^2$ depends strongly on the
functional form chosen for $\xi(\omega)$. Thus e.g. ARGUS from its data finds
\cite{ARG} values from 1.17 to 2.31 depending on the choice of the function.

We discuss the Isgur-Wise functions from the point of view of the (modified)
MIT bag model \cite{SAD}, but some of our conclusions are more general than the
model. Let us note first that the overlap should be a decreasing function of
$\omega$. Thus

\begin{equation}
\label{genbjo}
\xi(\omega) \leq \sqrt{\frac{2}{\omega+1}} = 1 - \frac{\omega-1}{4} +
O((\omega-1)^2)
\end{equation}

\noindent This bound has been pointed out by de Rafael and Taron \cite{DRT}.
The implication that $\rho^2 > \frac{1}{4}$ is known as the Bjorken limit
\cite{BJO}.

The calculations in the framework of the MIT bag model depend on the reference
frame chosen. We choose the (modified \cite{HOG}) Breit frame, where the
velocities of $H_b$ and $H_c$ are equal and opposite. The velocity of $H_c$
will be denoted $\vec{v}$ and its Lorentz factor $\gamma$. Let us note the
kinematical identities

\begin{equation}
\gamma = \sqrt{1+\vec{v}^2} = \sqrt{\frac{\omega+1}{2}}
\end{equation}

In the Breit frame the overlap function for mesons has the form

\begin{equation}
<\Phi_{Hcl}|\Phi_{Hbl}> = \int_{CB}\Phi^{\dag}(L^{-1}_{\vec{v}}(0,\vec{x}))
S^{\dag}(\vec{v})S(-\vec{v})\Phi(L^{-1}_{-\vec{v}}(0,\vec{x})) d^3x
\end{equation}

\noindent Here

\begin{equation}
\label{timsep}
\Phi(t,\vec{x}) = \Phi(\vec{x})exp(-iE_qt)
\end{equation}

\noindent is the wave function of the light component in its rest frame and
$E_q$ is the energy of the light quark. For the ground state to ground
state transitions, which we are considering, these functions for $H_b$ and
$H_c$ are identical, thus we omit the subscripts. Choosing the $z$-axis along
the vector $\vec{v}$, we have the Lorentz transformations

\begin {equation}
L^{-1}_{\pm\vec{v}}(0,\vec{x}) = (\mp vz, x, y, \gamma z)
\end{equation}

\noindent The wave functions relevant for the overlap integral are at $t=0$ in
the Breit frame and not at $t=0$ in the rest frames of the particles. For
$\vec{v} \neq
0$ this will introduce oscillating factors into the integrand and reduce the
overlap. The factors $S(\pm\vec{v})$ are Lorentz boosts. We only need to know
that they are real. Thus $S^{\dag}(\vec{v}) = S(\vec{v}) = S^{-1}(-\vec{v})$
and the two boost factors cancel. In the calculation of the normalization
integral for a bag moving with velocity $\vec{v}$, the corresponding factor
$S^{\dag}(\vec{v})S(\vec{v}) \neq 1$ cancels the effect of the Lorentz
contraction of the bag. The absence of this term gives a further reduction of
the overlap for $\omega \neq 1$. The integration is over the Lorentz -
contracted bag CB, which is an ellipsoid with semiaxes of length $R$ in the
directions perpendicular to the $z$-axis and of length $R/\gamma$ along the
$z$-axis. Rescaling the $z$-variable by a factor $\gamma$ we may rewrite the
overlap as an integral over a spherical bag $B$ of radius $R$

\begin{equation}
\label{overla}
<\Phi_{Hcl}|\Phi_{Hbl}> = \gamma^{-1}\int_B\Phi^{\dag}(\vec{x})\Phi(\vec{x})
e^{2iv\gamma^{-1}E_qz} d^3x
\end{equation}

A baryon differs from the meson in that it has two light quarks in the light
componenet. Therefore, its overlap function is just the square of the overlap
(\ref{overla}).

Since the integral in (\ref{overla}) is a decreasing function of $v$, we find
the inequality

\begin{equation}
\label{newbjo}
\xi(\omega) \leq \sqrt{\frac{2}{\omega+1}}^{N+1} = 1 -\frac{1+N}{4}(\omega-1)
+O((\omega-1)^2)
\end{equation}

\noindent where $N=1$ for mesons and $N=2$ for baryons. The corresponding
limits on $\rho^2$

\begin{equation}
\rho^2 \geq \frac{1+N}{4}
\end{equation}

\noindent are already of practical interest. E.g. Blok and Shifman \cite{BS}
from a particularly careful analysis of sum rules obtain for the $B$ meson
$\rho^2 = 0.65 \pm 0.15$, just above our limit.

In order to calculate $\rho^2$ from the full formula for the overlap, it is
enough to consider terms up to second order in velocity. Using the spherical
symmetry of the ground state one obtains

\begin{equation}
\rho^2 = \frac{N+1}{4} + \frac{1}{3}NE_q^2<r^2>
\end{equation}

In the bag model the product $E_q^2<r^2>$ is a slowly increasing function of
the dimensionless variable $m_qR$. We show it in Fig. 1. Using the model
it is possible to express the bag radius $R$ by the masses of the quarks
constituting the light component. The resulting dependence of $\rho^2$ on the
quark mass(es) is shown in Fig. 2 for three cases: for mesons the variable is
the mass of the light quark, for barions with two light quarks of equal mass
like $\Lambda_c$ or $\Omega_c$ the variable is the mass of either quark, for
baryons with one zero mass quark and one strange quark the variable is the mass
of the strange quark. Using the figure it is possible to find the bag model
prediction for $\rho^2$ corresponding to the ground state of any particle with
exactly one heavy quark. For the decays $\overline{B} \rightarrow
D^{(*)}l\overline{\nu}$ and $\Lambda _b \rightarrow \Lambda_c l \overline\nu$
we have $m_q=0$, which corresponds to

\begin{eqnarray}
\rho^2_B &=& 1.239\\
\rho^2_\Lambda &=& 2.228
\end{eqnarray}

For the decays $\overline{B}_s \rightarrow D_s^{(*)}l\overline{\nu}$ we use
\cite{SAD} $m_s = 0.291$, which yields

\begin{equation}
\rho^2_{B_s} = 1.625
\end{equation}

An exact evaluation of the overlap integral for finite $\omega-1$ yields a
result, which is very well approximated by the function deduced from a study of
sum rules by Neubert \cite{NEU}. The error of this approximation for $1 \leq
\omega \leq 3$ is less than 1 per cent. Thus, we use for $\overline{B}
\rightarrow D^{(*)}l\overline\nu$

\begin{equation}
\xi^{(B)}(\omega) = \left(\frac{2}{\omega+1}\right)^{2+\frac{0.6}{\omega}}
\end{equation}

\noindent and consequently for $\Lambda_b \rightarrow \Lambda_c l \overline\nu$

\begin{equation}
\xi^{\Lambda}(\omega) =
\left(\frac{2}{\omega+1}\right)^{3.5+\frac{1.2}{\omega}}
\end{equation}

\noindent For $\overline{B}_s \rightarrow D_s^{(*)}l\overline\nu$ we find
similarly

\begin{equation}
\xi^{B_s}(\omega) = \left(\frac{2}{\omega+1}\right)^{2.7+\frac{0.6}{\omega}}
\end{equation}

\noindent with an error below two per cent for $1 \leq \omega \leq 3$.

A comparison of our function $\xi^B(\omega)$ with the recent ARGUS data is show
in Fig. 3. As a by-product of this comparison we obtain the element of the
Cabibbo-Kobayashi-Masakawa matrix: $V_{cb} = 0.0420\sqrt{(\tau_{B}/1.29ps)}$.
The $\chi^2$ of this fit
is 7.0 for 7 degrees of freedom, which is very satisfactory. As a further check
we have calculated the branching ratios for the decays $\overline{B}
\rightarrow D^{(*)}l\overline\nu$. The results obtained using $V_{cb} = 0.043$
from \cite{PDG} are

\begin{eqnarray}
BR(\overline{B} \rightarrow D l\overline\nu) &=& 1.80(\tau_B/1.29ps)\% \\
BR(\overline{B} \rightarrow D^* l\overline\nu) &=& 5.46(\tau_B/1.29ps)\%
\end{eqnarray}

\noindent to be comapred with the experimental results \cite{PDG} (averaged
over the charges of the decaying $B$) $(1.7\pm0.4)\%$ and $(4.8\pm0.6)\%$.

We predict moreover

\begin{eqnarray}
BR(\overline{B}_s \rightarrow D_sl\overline\nu) &=&
1.51(\tau_{B_s}/1.29ps)\%\\
BR(\overline{B}_s \rightarrow D_s^{(*)}l\overline\nu) &=&
4.95(\tau_{B_s}/1.29ps)\%\\
BR(\Lambda_b \rightarrow \Lambda_c l\overline\nu) &=&
6.09(\tau_{\Lambda_b}/1.29ps)\%
\end{eqnarray}

\noindent For these decays one can as yet make no comparison with experiment.
Theoretical predictions, on the other hand, are ambiguous. Thus, Jenkins et al.
conclude from chiral perturbation theory that the ratio
$\xi^{B_s}(\omega)/\xi^B(\omega)$ should be an increasing function of $\omega$
\cite{JS}. The bag model gives the opposite prediction. The bag model predicts
that the function $\xi^\Lambda(\omega)$ decreses with increasing $\omega$ much
faster than the function $\xi^B(\omega)$. A similar prediction is made by
K\"onig et al \cite{K^4} from a calculation of overlaps in the infinite
momentum frame. On the other hand, models based on meson exchange
\cite{MS},\cite{SJR} suggest that the Isgur-Wise functions for the two cases
should be similar. A steeper Isgur-Wise function means that the ground state to
ground state semileptonic decays are a smaller fraction of all the semileptonic
decays of the parent particle. Experimental data should soon shed some light on
these questions.

Let us note finally that the function $\xi^B(\omega)$ can be also used to get a
crude estimate of the branching ratio for rare decays like $B \rightarrow
K^*\gamma$ \cite{AOM}. Our function is significantly smaller than the function
used in ref. \cite{AOM} in the region relevant for the $b \rightarrow c$
transitions. For the $B \rightarrow K^*$ transition, however, which corresponds
to $\omega \approx 3$, the two functions coincide. Thus we support the results
from ref. \cite{AOM}.

\section{Conclusions}

The (modified) MIT bag model gives plausible predictions for the Isgur-Wise
functions corresponding to the groud state to ground state transitions. Several
results seem more general than the model used:
\begin{itemize}
\item The parameter $\rho^2$ should satisfy inequality (\ref{newbjo}), which is
stronger than the Bjorken inequality \cite{BJO}.
\item The Isgur-Wise functions for baryons should be much steeper functions of
$\omega$ than the corresponding functions for mesons. For corresponding mesons
and baryons the relation between the parameters $\rho^2$ is $\rho^2_{baryon} =
2\rho^2_{meson} - 0.25$.
\item The Isgur-Wise function for the decays $\overline{B}_s \rightarrow
D_s^{(*)}l\overline\nu$ should be somewhat steeper than those for the decays
$\overline{B} \rightarrow D^{(*)}l\overline\nu$. With the quark mass increasing
from about zero to $m_s$, the slight decrease of the radius pointed out in
\cite{JS} is in our approach more than compensated by the increase of the quark
energy.
\end{itemize}
Also the observation that the Isgur-Wise function for $\overline{B} \rightarrow
D^{(*)}l\overline\nu$ from the present analysis is almost identical with the
function obtained by Neubert \cite{NEU} from sum rules, may mean that our
results are more general than the model used to derive them.

The calculation is the most reliable for small recoil velocities, but we find
plausible results for all recoils of interest. This is analogous to the
analysis in the framework of the nonrelativistic quark model as presented in
ref. \cite{ISGW}. There, however, the Isgur-Wise functions obtained seem to be
too flat. The original result $\rho^2 \approx 0.3$ is multiplied by two in
order to get a more reasonable $\omega$ dependence \cite{ISGW}. We explain this
fact as a result of two factors: In the nonrelativistic calculation the boosts,
which contribute 0.25 to $\rho^2$, are absent. Moreover, the quark energy $E_q$
is replaced by the constituent quark mass, which is smaller.

The overall agreement of our results with the little data available is good,
but there are hints that we may be overestimating the branching ratio for the
decay $\overline{B} \rightarrow D^* l\overline\nu$. If experiment supports
this, it would be a confirmation of the analysis of Amundson and Rosner
\cite{AR}, who find that the $O(1/m_Q)$ and the QCD corrections
cancel for the decay $\overline{B} \rightarrow D l \overline\nu$, while the
negative QCD correction dominates for
 $\overline{B} \rightarrow D^* l \overline\nu$.

\end {document}